\documentclass[]{aa}  

\usepackage{color, graphicx}
\usepackage{txfonts}
\begin{document}

   \title{Sectoral r modes and periodic RV variations of Sun-like stars}


   \author{A.~F.~Lanza
          \inst{1}
          \and
          L.~Gizon\inst{2,3,4}
          \and
          T.~V.~Zaqarashvili\inst{5,6,7}%
          \and Z.-C. Liang\inst{2}
          \and K.~Rodenbeck\inst{3,2}
          }

   \institute{INAF-Osservatorio Astrofisico di Catania, Via S.~Sofia, 78 -- 95123 Catania, Italy\\
              \email{antonino.lanza@inaf.it}
         \and
    Max-Planck-Institut f\"ur Sonnensystemforschung, Justus-von-Liebig-Weg 3, 37077 G\"ottingen, Germany\\
    \email{gizon@mps.mpg.de, zhichao@mps.mpg.de}
    \and
     Georg-August-Universit\"at, Institut f\"ur Astrophysik, Friedrich-Hund-Platz 1, 37077 G\"ottingen, Germany\\
     \email{rodenbeck@mps.mpg.de}
     \and 
     Center for Space Science, NYUAD Institute, New York University Abu Dhabi, PO Box 129188, Abu Dhabi, UAE        
       \and 
     Space Research Institute, Austrian Academy of Sciences, Schmiedlstra{\ss}e 6, 8042 Graz, Austria\\
      \email{Teimuraz.Zaqarashvili@oeaw.ac.at}
     \and 
     Abastumani Astrophysical Observatory at Ilia State University, 3/5 Cholokashvili Avenue, 0162, Tbilisi, Georgia
     \and 
     IGAM-Kanzelh\"ohe Observatory, Institute of Physics, University of Graz, Universit\"atsplatz 5, 8010 Graz, Austria    
             }

   \date{Received ... ; accepted ...}

 
  \abstract
   {Radial velocity (hereafter RV) measurements are used to search for planets orbiting late-type main-sequence stars and confirm the transiting planets.}
   {The most advanced spectrometers are now approaching a precision of $\sim 10$~cm/s that implies the need to identify and correct for all possible sources of RV oscillations intrinsic to the star down to this level and possibly beyond. The recent discovery of global-scale equatorial Rossby waves in the Sun, also called r~modes, prompted us to investigate their possible signature in stellar RV measurements.  R~modes are toroidal modes of oscillation whose restoring force is the Coriolis force; they propagate in the retrograde direction in a frame that corotates with the star. The solar r~modes with azimuthal orders $3 \leq m \lesssim 15$ were identified unambiguously because of their dispersion relation and their long e-folding lifetimes of hundreds of days.}
   {In this paper, we simulate the RV oscillations produced by sectoral r~modes with $2 \leq m \leq 5$, by assuming a stellar rotation period of $25.54$ days and a maximum amplitude of the surface velocity of each mode of 2~m/s.
   This amplitude is representative of the solar measurements, except for the $m=2$ mode which has not yet been observed on the Sun.  }
   {Sectoral r~modes with  azimuthal orders $m=2$ and $3$ would produce RV oscillations with amplitudes of $76.4$ and $19.6$~cm/s and periods of $19.16$ and $10.22$~days, respectively, for a star with an inclination of the rotation axis to the line of sight  $i=60^{\circ}$.  Therefore, they may produce rather sharp peaks in the Fourier spectrum of the radial velocity time series that could lead to spurious planetary detections.  
   }
   {Sectoral r~modes may represent a source of confusion in the case of slowly rotating inactive stars that are preferential targets for RV planet search. The main limitation of the present investigation is the lack of observational constraint on the amplitude of the $m=2$ mode on the Sun. }


   \maketitle
%

\section{Introduction}
\label{intro}
The search for Earth-like planets around late-type stars is one of the most active areas of modern astronomy. The radial velocity (hereafter RV) oscillations of the host stars induced by the motion of orbiting planets have been used to discover the first exoplanet and many others to date \citep[e.g.,][]{Mayoretal14}. Moreover, to confirm candidates discovered by means of transits and measure their masses, it is necessary to measure the radial velocity (hereafter RV) oscillations of the host stars. This RV follow up often leads to the discovery of additional non-transiting planets around the host stars, allowing a better characterization of the architectures of their planetary systems. 

The advent of HARPS \citep{Mayoretal03} and HARPS-N \citep{Cosentinoetal12} has pushed the RV measurement precision and long-term stability below 1 m/s on bright stars ($V \la 8-10$), while the recently commissioned ESPRESSO spectrograph aims at a precision of 0.1 m/s \citep{Pepeetal14} making it possible to detect the wobble of a planet like our Earth on a one-year orbit around a sun-like star, provided that stellar intrinsic RV variations can be  properly corrected. The amplitude of the Sun RV oscillation induced by the Earth motion is $\sim 9.2$~cm/s, while amplitudes of the order of 2-4 m/s are observed in M~dwarfs orbited by Earth-mass planets in their habitable zones  \citep[e.g.,][]{Afferetal16}. 

The critical limitation in discovering Earth-like planets through the RV method is the  intrinsic variability of their hosts that is produced by several physical processes. Photospheric granulation and p-mode oscillations can be averaged out by designing the observations in an appropriate way \citep{Dumusqueetal11}. Photospheric magnetic fields produce brightness inhomogeneities akin sunspots and faculae that perturb the flux balance between the redshifted and the blueshifted portions of the disc of a rotating star leading to a systematic distortion of their spectral line profiles that in turn induces an apparent variation of their RV. Moreover, surface magnetic fields locally quench convective motions responsible for the apparent spectral line blueshifts, thus producing an apparent increase in the disc-integrated RV measurements \citep[e.g.,][]{SaarDonahue97,Meunieretal10}. Such effects are modulated by the stellar rotation and the evolution of the photospheric magnetic fields in active regions generally showing a periodicity  close to the stellar rotation period and its first two or three harmonics \citep{Boisseetal11}. By correlating the RV variations with the chromospheric activity index $\log R^{\prime}_{\rm HK}$ \citep[e.g.,][]{Lovisetal11,Lanzaetal16} or with suitable indicators of the line profile distortions \citep[e.g.][]{Dumusqueetal14,Lanzaetal18}, such intrinsic RV variations can be identified and corrected to some extent. 
However, there are other phenomena that can induce an apparent quasi-periodic RV variation in sun-like stars that are more difficult to recognize because they are not directly associated with variations of activity indicators, for example a modulation of the meridional circulation as proposed by \citet{Makarov10}. 
In the present work, we propose that r-mode oscillations recently detected in the Sun \citep{Loptienetal18} can be another phenomenon of this kind, which could potentially lead to false planetary detections in late-type stars. 

\section{Observations}
\label{observations}
The so-called r modes are the toroidal modes of  oscillation of a rotating star \citep[see, e.g.,][]{Papaloizou78, Saio82}, for which the dominant restoring force is the Coriolis force. They are associated with a pattern of radial vorticity of alternating sign that propagates in the retrograde direction in a reference frame that is co-rotating with the star. 

Recently, \citet{Loptienetal18} mapped the horizontal velocity field  in the solar photosphere for the period 2010--2016 by using granules observed in intensity by HMI/SDO as tracers of the flow. By analyzing the sectoral power spectra of the radial component of vorticity, $\zeta$, they discovered oscillations with frequencies corresponding to the classical sectoral Rossby waves. By construction, sectoral power spectra are obtained by projecting the spatial data onto spherical harmonics with equal angular degree $l$ and azimuthal wavenumber $m$. Due to the  `shrinking-Sun effect' \citep{Loptienetal16}, the horizontal velocity maps obtained by granulation tracking could not be used to characterize modes with $m \leq 2$.  The non-sectoral power spectra did not reveal any additional modes of oscillation. 

The detected r modes were identified through their frequencies in the co-rotating frame (see Sect.~\ref{model}). The rms radial vorticity of the modes ranges between $0.75 \times 10^{-8}$~s$^{-1}$ and $2.42 \times 10^{-8}$~s$^{-1}$. \citet{Loptienetal18} noted that the $m=3$ sectoral mode displayed a double-peak structure in the power spectrum for 2010--2016, so the mode parameters were not studied in detail in this case.  The e-folding lifetime of the vorticity pattern associated with individual solar r~modes is up to eleven solar rotation periods, as indicated by the high quality factors of the peaks in the power spectrum.

\citet{Loptienetal18} also confirmed the r-mode detection in the shallow subsurface layers using ring-diagram helioseismology.  A recent time-distance helioseismology analysis by \citet{Liangetal18} extends these results to deeper layers and a time interval of 21 yr (including data from SOHO/MDI). They confirm e-folding lifetimes of almost two years for the sectoral $m=3$ and $m=5$ modes and an upper limit for the maximum surface velocity of the undetected $m=2$ mode of about 0.7~m/s. \citet{HanasogeMandal19} also confirmed the detection of sectoral modes with odd values of $m$, the only ones their specific normal-mode coupling method is sensitive to.  

Evidence of r-mode oscillations in stars other than the Sun was obtained by, e.g., \citet{VanReeth16} and \citet{Saioetal18}  by analysing photometric timeseries of the Kepler space telescope. These stars are early-type B and A main-sequence stars, $\gamma$ Doradus variables, or stars in eccentric close binaries with strong tidal interactions, i.e. stars that are not usual targets for planet search through the RV method because of their rapid rotation.  The relative amplitudes of the light variations associated with r modes in these stars are of the order of $\delta L/L \sim 10^{-3}-10^{-4}$. The horizontal velocity of their oscillations is of the order of $ v_{\rm h} \sim \sigma (\delta L/L)\, GM R^{-2}\Omega^{-2}$, where $\sigma$ is the frequency of the modes, $G$ the gravitation constant, $M$ the mass of the star, $R$ its radius, and $\Omega$ its rotation frequency \citep[cf., e.g.,][]{Provostetal81,Saio82,Kepler84}. Therefore, they have horizontal velocities of the order of km/s, that is at least a factor of $10^{3}$ larger than in the case of the Sun.

\section{Model}
\label{model}
The surface displacement $\vec \xi$ associated with an r mode is dominated by the toroidal component of displacement \citep[see, e.g.,][Sect. IIb]{Kepler84}:
\begin{equation}
{\vec \xi} \simeq \text{Re}  \left\{ {\vec T}\right\}, 
\label{eq1}
\end{equation}
where, for a mode with  spherical harmonic degree $l$ and azimuthal order $m$, we have 
\begin{equation}
{\vec T} = K_{lm} \left( \mathbf{\hat{\Theta}} \frac{1}{\sin \Theta} \frac{\partial}{\partial \Phi}  -   \mathbf{\hat{\Phi}} \frac{\partial}{\partial \Theta} \right) Y_l^m (\Theta, \Phi) \exp (-i \sigma_{\rm obs} t) .
\label{eq2}
\end{equation}  
In the above expression $\Theta$ is the colatitude measured from the stellar rotation axis,   $\Phi$  the longitude measured in the inertial frame of the observer, $Y_l^m$ is a spherical harmonic function, and $K_{lm}$ fixes the amplitude of the  mode. 
The frequency of oscillation in the inertial frame, $\sigma_{\rm obs}$,  is related to the stellar angular velocity  $\Omega$ and the mode frequency $\sigma$ in the co-rotating frame through
\begin{equation}
\sigma_{\rm obs} = m \Omega + \sigma,
\label{eq4}
\end{equation} 
where, in the limit of slow uniform rotation ($\Omega^{2} \ll GMR^{-3}$) \citep[cf.][]{Saioetal18}, 
\begin{equation}
\sigma = -\frac{2 m \Omega}{l(l+1)}.
\label{eq5}
\end{equation}
In the inertial reference frame of a distant observer, the dipole sectoral mode  has a vanishing frequency, so we  consider only the effects of sectoral modes with $l=m \geq 2$ on the RV of the star. 

The velocity associated with an r-mode with given $l$, $m$, and prescribed parameters is the total time derivative of $\vec \xi$; its  explicit expression can be found in Sect.~III of \citet{Kepler84}. We compute the radial velocity of each surface element by means of  Eq.~(68) of \citet{Kepler84} assuming an inclination angle  $i$ for the stellar rotation axis to the line of sight. For simplicity, we assume a star with the same radius as the Sun, an angular velocity $\Omega = 2.84738$~$\mu$rad~s$^{-1}$, corresponding to the solar equatorial rotation period of $25.54$ days, and a linear limb-darkening coefficient $u=0.684$, appropriate for the continuum of the V passband \citep[cf.][Table~1]{Diaz-Cordovesetal95}.

To compute the RV of the star, we consider an average spectral line whose profile is obtained by integrating over the disc of the star the local line profiles emerging from the individual surface elements \citep[cf.][]{Lanzaetal11}. The local line profile is assumed to be a Gaussian displaced by the local radial velocity computed as explained above. We include the effect of the limb-darkening variation in the continuum of the spectral lines, but other effects, such as the possible variation of the line absorption coefficient with the position on the disc of the star or the effects of temperature inhomogeneities in active regions, are not included for simplicity. 

The RV of a star is generally measured by computing the cross-correlation function (CCF) between its spectrum and a binary line mask depending on its spectral type. The CCF is then fitted with a Gaussian and its central wavelength is used as a measure of the RV of the star \citep[e.g.,][]{Baranneetal96}. In our case, we derive the RV by considering in place of the CCF the average spectral line profile computed above. Other methods for measuring the RV have been proposed by, e.g., \citet{Gallandetal05},  \citet{AngladaEscudeButler12}, or \citet{Astudillo-Defuretal17} and have been applied to stars of spectral type A or F, having a relatively low number of spectral lines in their optical spectra and relatively fast rotation, or to M-type dwarfs, respectively.  

A crucial parameter in the computation of our synthetic RV oscillations is the amplitude $K_{mm}$ of the individual sectoral r~modes. 
Guided by the solar observations  \citep{Loptienetal18}, we fix $K_{mm}$ in order to have a maximum surface velocity of 2~m/s for each mode with $m \leq 5$. We do not consider~modes with $l > m$, which are not observed in the Sun. 

A simplification of our approach is the neglect of the effects of differential rotation on r~modes. As they are retrograde modes, their phase speed with respect to the local plasma will change with latitude as a consequence of the latitudinal differential rotation \citep{Wolff98}. 

The amplitude $K_{mm}$ may be larger in stars with a rotation rate and/or a spectral type different from those of the Sun. To make predictions of the r-mode amplitudes in late-type stars, the mechanisms of excitation and damping would have to be identified, but unfortunately we can presently make only conjectures about them. If r~modes are excited by convection, we expect that stars with a convective flow faster than the Sun, such as main sequence F-type stars, may display larger amplitudes. If they are excited by the shear associated with differential rotation \citep[][]{Saioetal18}, their amplitude may depend on the amplitude of the differential rotation both in the radial and in the latitudinal directions. 

\setcounter{table}{0}
\section{Results}
\label{results}
\begin{figure}
   \resizebox{\hsize}{!}{\includegraphics{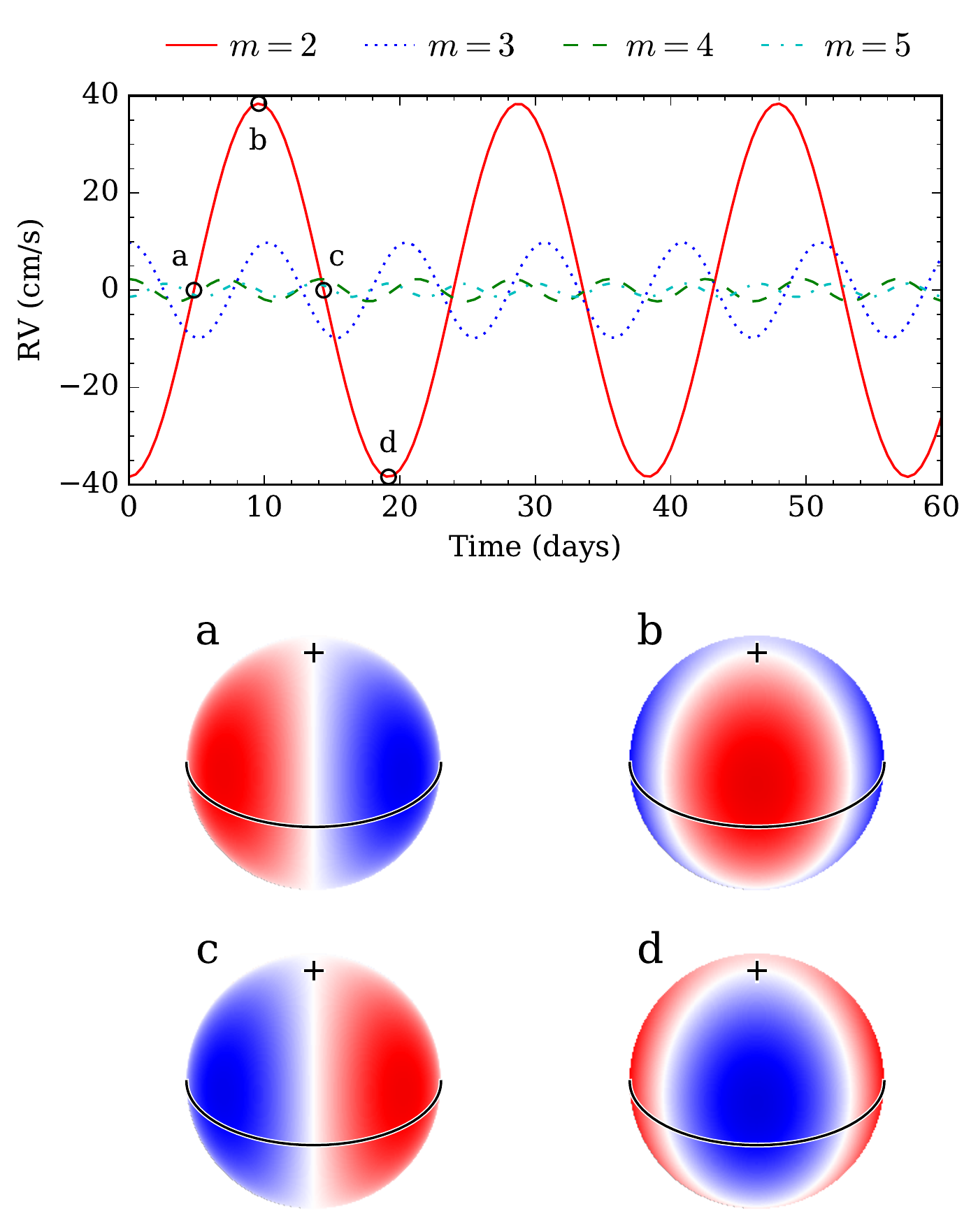}}
      \caption{ \label{FigVibStab}
      \emph{Top}: Simulated RV variations of a sun-like star produced by individual r~modes as seen by a distant observer as a function of the time. The spin axis of the star has an inclination of $i=60^{\circ}$ to the line of sight. Different linestyles refer to different sectoral modes: the solid line is for $m=2$ mode, the dotted line for $m=3$, the dashed line for $m=4$, and the dash-dotted line for $m=5$. See Table~1 
      for the amplitude,  the period, and the rms radial vorticity of the modes.
      \emph{Bottom}: The line-of-sight velocity on the disk for $m=2$ at four different phases a--d that are marked with circles in the top panel.
      The color scale saturates at $\pm2$~m/s: blue is toward and red is away from the observer.
      The plus symbols indicate the north pole and the black lines the equator.
      }
   \end{figure}

In Fig.~\ref{FigVibStab} we plot the RV oscillations for a sun-like star with sectoral r modes the amplitudes of which have been fixed to have a maximum surface velocity $\max\{v_{\rm surf}\} =2$~m/s for each mode. The surface velocity is defined as $v_{\rm surf} = (v_{\Theta}^{2} + v_{\Phi}^{2})^{1/2}$, where $v_{\Theta}$ is the meridional and $v_{\Phi}$ the azimuthal component of the velocity (cf. Sect.~\ref{model}). The amplitude (from minimum to maximum velocity) of the oscillations $A_{0}$ is given in Table~1 together with their period and rms radial vorticity $\zeta$ assuming that the star is rotating with the  solar rotation period of $25.54$ days and considering an inclination $i=30^{\circ}$ or $i=60^{\circ}$. For an inclination $i=0^{\circ}$ or $i=90^{\circ}$ the amplitudes vanish because of the symmetry leading to a perfect cancellation of the integrated radial velocity. The rms radial vorticity is computed considering the velocity field in the $\pm\, 20^{\circ}$ latitude range to allow a direct comparison with the observations by \citet{Loptienetal18}. We assume here that the amplitude of the modes is not damped, which is a rather good approximation given the long lifetimes of the solar r~modes (cf. Sect.~\ref{observations}).   With our choice of the $K_{mm}$'s,  the rms vorticities of the simulated r~modes are  smaller by a factor of $\approx 1.5-2$  than those observed by \citet{Loptienetal18} for $l=m \geq 4$, implying that our choice is conservative. 
Although $m=2$ has not been observed and $m=3$ is rather weak in comparison with $m \geq 4$ in the solar case, the same magnitude is assumed for all the modes implemented in this work.  This allows us to immediately see the increasing cancellation effects reducing the RV amplitudes of the modes with higher $m$ when their variations are integrated over the stellar disc. Because the observed RV amplitude and rms vorticity scale in direct proportion to the maximum surface velocity for a given mode, it is simple to compute its RV amplitude for different amplitudes of the surface velocity (see an example below). Indeed, given our ignorance of the excitation and damping mechanisms of these modes in the Sun and late-type stars, we cannot rely on the theory to predict their amplitudes.

%
   \begin{figure}
   \hspace*{-1cm}
   \resizebox{1.3\hsize}{!}{\includegraphics[angle=90,width=11cm,height=10cm]{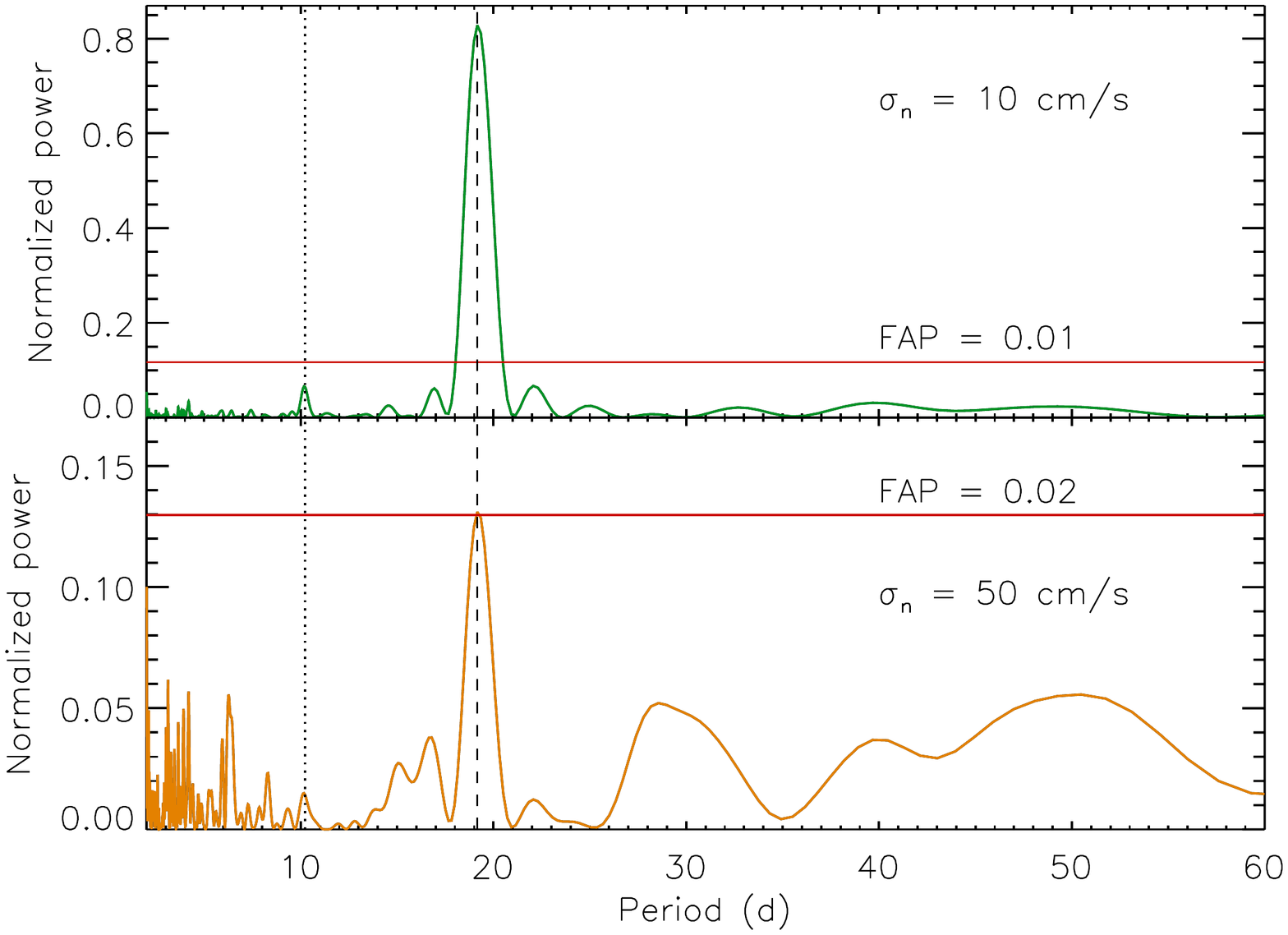}}
   \vspace*{-1.3cm}
      \caption{ \label{periodogram_star}
      \emph{Top}: Periodogram of the simulated radial velocity timeseries of a sun-like star  showing sectoral r~modes with $m=2$ and $3$ as seen with an inclination $i=60^{\circ}$. The RV amplitudes and the periods of the modes are given in Table~1 and corresponds to a maximum horizontal velocity of the modes of 2~m/s at the stellar surface. The periodogram is normalized according to the definition of the GLS given by \citet{ZechmeisterKuerster09}. A normally distributed random noise with a standard deviation $\sigma_{\rm n}$ of 10~cm/s was added. The vertical dashed line marks the period of the $m=2$ mode and the dotted line that of the $m=3$ mode. The red horizonthal line marks the level of 1 percent false-alarm Probability (FAP) as determined by analysing $10\,000$ random shufflings of the original time series. 
            \emph{Bottom}: Same as the upper panel, but for a standard deviation of the noise $\sigma_{\rm n}$ of 50~cm/s. Now the red horizonthal line marks the level corresponding to a false-alarm probability of 2 percent. 
      }
   \end{figure}
%
   \begin{figure}
   \hspace*{-1cm}
   \resizebox{1.3\hsize}{!}{\includegraphics[angle=90,width=11cm,height=10cm]{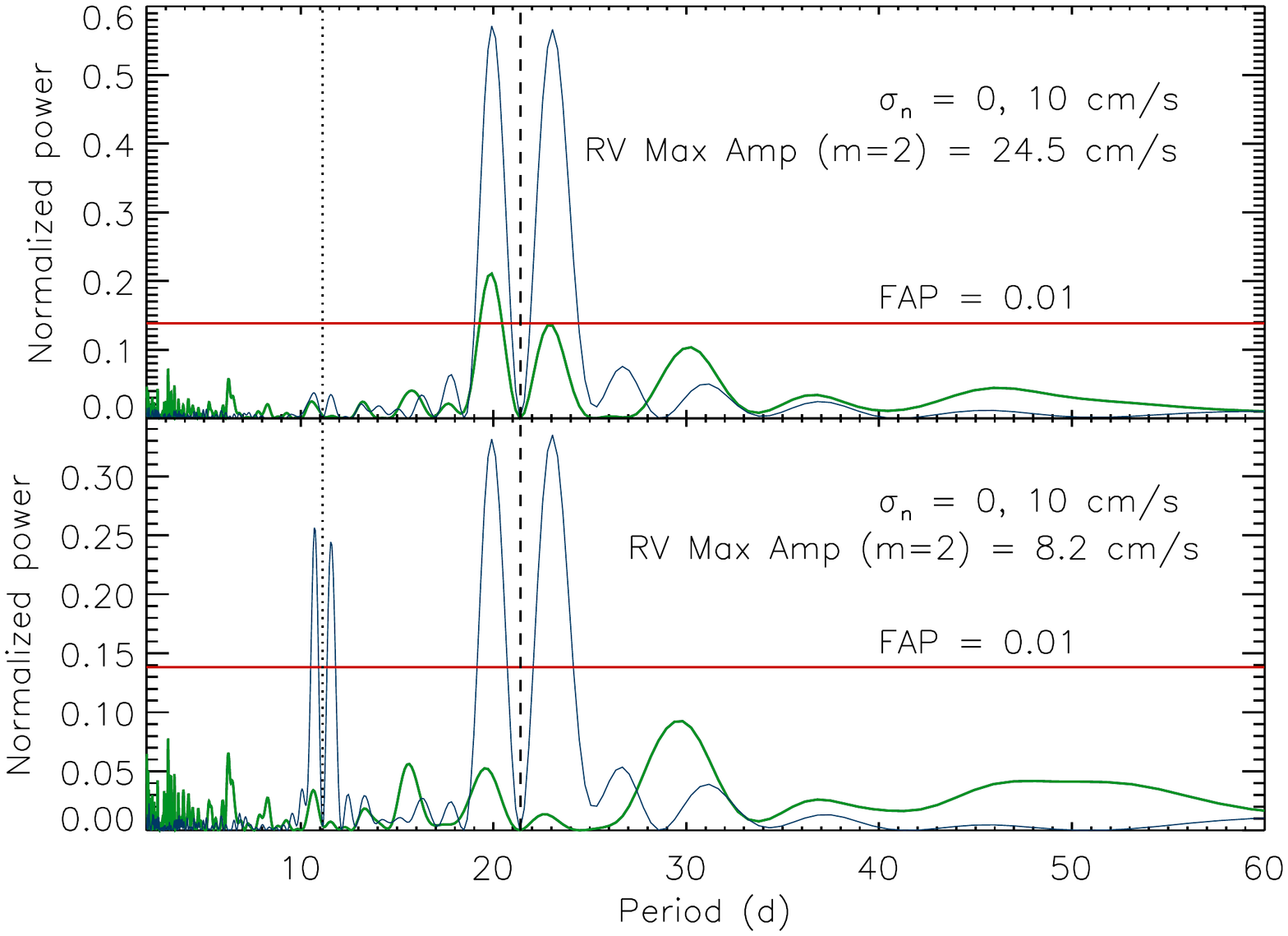}}
   \vspace*{-1.3cm}
      \caption{ \label{periodogram_sun}
      \emph{Top}: Periodogram of the simulated radial velocity timeseries of the Sun as a star with sectoral r~modes with $m=2$ and $m=3$ as seen by an Earth observer. The RV maximum amplitude of the $m=2$ mode is  24.5~cm/s, while the amplitude of the $m=3$ mode is 7.1~cm/s;  the periods of the modes are given in the text. These RV amplitudes correspond to a maximum horizontal velocity of the modes of 2~m/s at the surface of the Sun. The periodogram of a noiseless time series  is plotted with a blue thinner solid line, while the periodogram obtained adding  a normally distributed random noise with a standard deviation $\sigma_{\rm n}=10$~cm/s is plotted with the thicker green solid line. The vertical dashed line marks the period of the $m=2$ mode and the dotted line that of the $m=3$ mode, if the inclination $B_{0}$ of the solar equator is fixed.  The red horizonthal line marks the level of 1 percent false-alarm probability (FAP). 
            \emph{Bottom}: Same as the upper panel, but the maximum RV amplitude of the sectoral $m=2$ mode is  assumed now $8.2$~cm/s, corresponding to the upper limit for the amplitude of this mode as found by \citet{Liangetal18}, while the amplitude of the $m=3$ mode is still 7.1~cm/s. The upper limit for the $m=2$ mode corresponds to a maximum horizontal velocity of 0.7~m/s at the surface of the Sun.    }
   \end{figure}
%
\begin{table}
\begin{center}
\caption{Peak-to-peak amplitudes, $A_{0}$, and sidereal oscillation periods of the RV variations produced by sectoral  r~modes in a Sun-like star inclined by  $i=30^{\circ}$ and $i=60^{\circ}$. All modes are assumed to have the same maximum surface horizontal velocity of 2 m/s.}
\begin{tabular}{crrrr}
\hline
mode & $A_{0}(i=30^{\circ})$ & $A_{0}(i=60^{\circ})$ & period & $\zeta$ rms\\
$(l,m)$ & (cm/s) & (cm/s) & (days) & $ (10^{-8}$ s$^{-1}$)\\
\hline
$(2,2)$ & 44.3 & 76.4& 19.16 & 0.24\\
$(3,3)$ &  6.7 & 19.5 & 10.22 & 0.38\\
$(4,4)$ & 0.6 & 4.4 &  7.10 & 0.53\\
$(5,5)$ & $\sim 10^{-4}$ & 2.3 & 5.47 & 0.68\\ 
\hline 
\end{tabular}
\end{center}
\label{table_1}
\end{table}
In our simulations, the sectoral modes giving the largest RV oscillations are $m=2$, followed by $m=3$ (cf. Table~1). If the maximum velocity of the modes is increased, for example, to 4~m/s, the amplitude of the RV oscillation produced by the $(2,2)$ mode becomes $153$~cm/s for $i=60^{\circ}$. The amplitude of the RV oscillations is rapidly decreasing with increasing $m$ due to increasing cancellation of the velocity patterns when  integrating over the disc of the star. For this reason, we consider only r~modes with $l \leq 5$. Although their frequencies are smaller than the rotation frequency in the co-rotating frame, in the observer's frame they have positive frequencies  $\sigma_{\rm obs}= [m-2/(m+1)]\Omega$,  tending to $m\Omega$ for large $m$ (cf. Eqs.~\ref{eq4} and~\ref{eq5}). 

In Fig.~\ref{periodogram_star}, we plot the periodogram of a simulated RV time series of 200 days with daily cadence and randomly missing observations amounting to 30 percent of the time to include the effect of  bad weather. We consider only the contributions of the sectoral modes with $m=2$ and 3 assumed to have a constant amplitude in a sun-like star viewed with an inclination of $60^{\circ}$ and add a normally distributed random noise with standard deviation $\sigma_{\rm n}$ of 10 or 50~cm/s. The periodogram is computed with the Generalized Lomb-Scargle (GLS) formalism of \citet{ZechmeisterKuerster09}.  The red horizontal lines at a normalized  power of 0.138 mark the level corresponding to a false-alarm probability (FAP) of 1~percent as derived by analysing $10\,000$ random shufflings of the time series. The level corresponding to a FAP of 5~percent (not marked) is 0.117. Therefore, the $m=2$ mode is detected with high significance, while the FAP of the $m=3$ mode is above 5~percent when we consider a noise with $\sigma_{\rm n}=10$~cm/s. With a standard deviation of the noise $\sigma_{\rm n}=50$~cm/s, the false-alarm-probability of the $m=2$ mode  is 0.019, indicating that a higher noise level may hamper a clear detection of this r mode. Note that, even if they are not significantly detected, the effects of the modulations due to the $m=2, 3$ modes remain hidden in the power spectrum and contribute to the intrinsic stellar RV variability at the 10-cm/s level. 

We repeated the above calculations for our Sun for which the apparent latitude $B_{0}$ of the centre of the disc oscillates between $-7^{\circ}.23$ and $+7^{\circ}.23$ over a period of one year. Assuming again $\max\{v_{\rm surf}\} = 2$~m/s, we find that the amplitude of the $m=2$ mode is modulated between zero and a maximum of $24.5$~cm/s. For the $m=3$ mode, we find a maximum amplitude of $7.1$~cm/s.

Because we observe the Sun from the Earth vantage point and because of the modulation of the amplitude due to the annual variation of the apparent inclination of the solar rotation axis ($B_0$ angle variations), the RV power of the $m=2$ mode would be split into two peaks with periods of $20.22$ and $22.74$~days, respectively \citep[cf.][]{Wolff86}. The RV power of the $m=3$ mode would be split into two equal peaks with periods of $10.82$ and $11.5$~days, respectively.

In Fig.~\ref{periodogram_sun}, we plot the GLS periodogram of a simulated daily RV time series of the Sun as a star with a total extension of 200 days and missing observations covering 30 percent of the time. We consider two possible RV amplitudes for the sectoral $m=2$ mode, that is, 24.5 or 8.1~cm/s, the latter corresponding to the upper limit of 0.7~m/s for $v_{\rm surf}$ as found by \citet{Liangetal18}, and simulate both noiseless time series and time series with a random normal noise of standard deviation of 10 cm/s. The red horizontal lines in the figure panels indicate the level of the normalized power corresponding to a false-alarm probability of 1~percent.  We conclude that even a modest amount of noise can hamper a significant detection of the r modes in the disc-integrated solar RV observations. Note the splitting of the power of the r modes into two close peaks having periods shorter and longer than the periods predicted in the case of a fixed $B_{0}$ angle, respectively, as a consequence of the yearly modulation of $B_{0}$ for an observer on the Earth.  

The peaks of the $m=3$ mode for the noiseless time series ($\sigma_{\rm n}=0$) are higher in the lower panel of Fig.~\ref{periodogram_sun} because the variance of that time series is lower than the variance of the time series in the upper panel, thanks to the smaller amplitude of the $m=2$ mode. Since the periodogram power is normalized to the weighted variance of the data around their mean \citep[cf. Eq.~(4) in][]{ZechmeisterKuerster09}, the normalized power of the signal of the $m=3$ mode increases. On the other hand, the level of 1~percent FAP for the time series with $\sigma_{\rm n}=10$~cm/s,  is not modified because it is determined by considering the distribution of the maximum values of the GLS of $10\,000$ randomly shuffled time series that have no periodic signals, but have the same variance of the original time series which leads to the same normalized level.

\section{Discussion}
\label{discussion}
Due to their long lifetimes, the sectoral r~modes are candidates to produce RV oscillations that may be misinterpreted as due to an orbiting Earth-mass planet.
According to the above simulations, the sectoral $m=2$ modes may have the highest amplitudes in RV measurements.
Unfortunately, we currently have no observational constraints on the surface amplitude of these modes on the Sun. 
The $m=2$ modes may or may not be excited to sufficiently large amplitudes to be detectable on Sun-like stars.
Additional information from solar observations and theory will be required to estimate mode amplitudes. We note that it is not excluded that quadrupole r modes could be excited to significant amplitudes through tidal interaction with a close-in planet \citep{Ogilvie14}.

The frequencies of the RV oscillations of r~modes in a star are close but not equal to the rotation rate $\Omega$, specifically they are $4\Omega/3$ and $5\Omega/2$  for the sectoral modes with $m=2$ and $3$. In the case of a star rotating like the Sun, this implies putative orbital periods of $19.16$ and $10.22$ days that would correspond to planets with minimum masses of $0.22$ and $0.07$ Earth mass, respectively \citep[see Table~1 and][]{Wright17}.  Although some cases of RV signals with a period ratio close to 4/3 have been reported \citep[e.g.,][]{Santosetal14}, other effects can contribute to produce peaks with the frequencies indicated above, notably spot intrinsic evolution, differential rotation \citep[e.g.][]{ReinholdGizon15} or the combination of differential rotation with spots on opposite hemispheres of a star \citep[e.g.][]{CollierCameronetal09}. Therefore, we expect the power spectrum of real active stars to be rather complex with the signals of r modes not unambiguously detectable in all the cases, even if their amplitudes are sufficiently large. 

The peak in the Fourier spectrum of the RV produced by an r~mode may persist for long time intervals because modes are long lived. Given a quality factor $Q \sim 10$, this would produce a peak very similar to those generally attributed to the reflex motion induced by an orbiting planet unless a long time-series with frequent sampling and covering several seasons is acquired to reveal the expected phase changes associated with the attenuation and re-excitation of the mode. 

Most of the active regions in slowly rotating stars such as the Sun  have lifetimes shorter than the rotation period. Only long-lived active regions and photospheric faculae  give a truly periodic signal in RV series. Episodically, large complexes of activity may last for 5-15 rotation periods \citep{Castenmilleretal86}. The associated RV periodicities are generally at the rotation period of the star and its first two or three harmonics \citep[cf.][]{Boisseetal11}, while the periodicities induced by r~modes would appear at different periods that could be misinterpreted as the periods of orbiting low-mass planets. 

Some stars that are targeted for planet search have rotation periods significantly longer than that of the Sun, up to $120-150$ days for some low-activity M-type stars \citep{Robertsonetal14}. These targets may have sectoral modes producing RV oscillations with periods of several tens of days that could be misinterpreted as planets in their habitable zones (such habitable zones are closer to the stars than in the case of the Sun because of the lower stellar luminosities).

\begin{figure}[t]
      \centering
\includegraphics[width=0.5\textwidth]{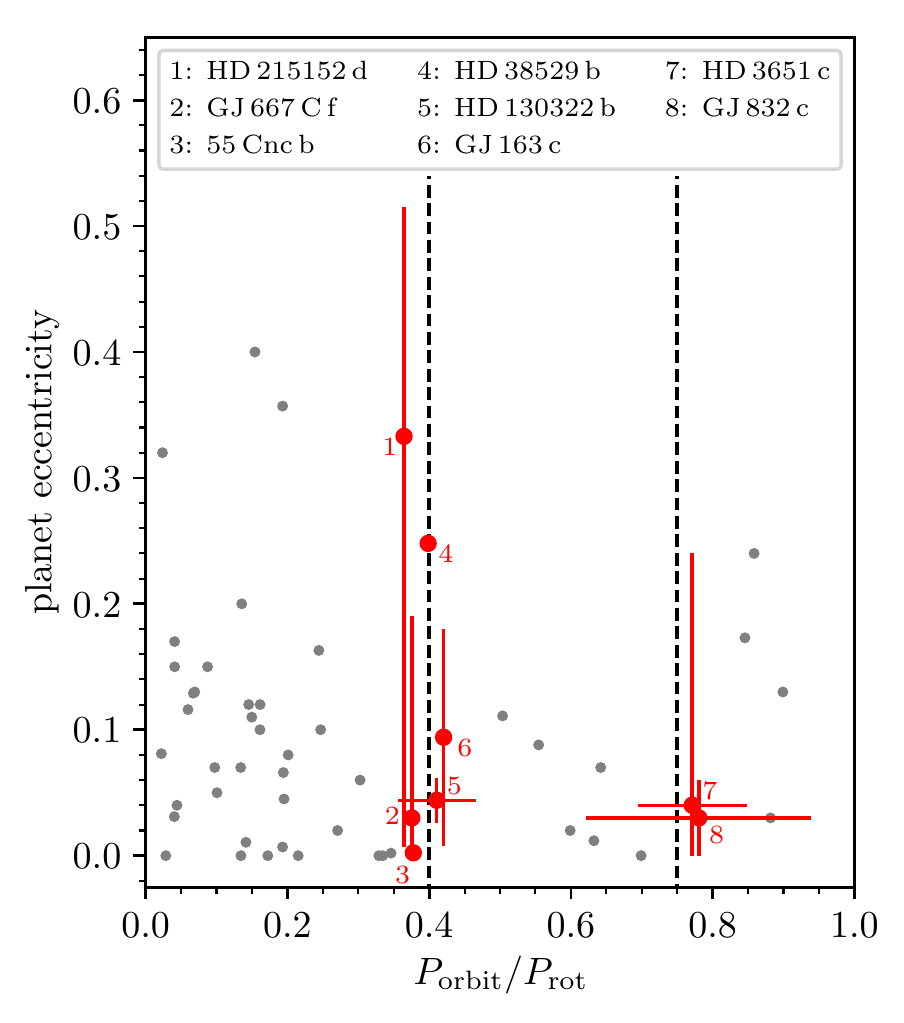}
      \caption{Ratios of exoplanet orbital periods ($P_{\rm orbit}$) to the rotation periods of their host stars ($P_{\rm rot}$), for exoplanets that have been observed in radial velocity only.  The exoplanet eccentricity is given on the vertical axis. The dashed lines indicate the expected ratios $P_{\rm orbit}/P_{\rm rot}$ for the $m=2$ and $m=3$ sectoral r~modes of stellar oscillation. Exoplanets with ratios close to these values are potential false positives and are labelled with a number listed in the top table and plotted in red with error bars.}
\label{fig:porb_prot}
  \end{figure}

As an application, we searched for the orbital periods of "confirmed exoplanets" that have been observed only in radial velocity from \url{exoplanet.eu}, excluding transiting exoplanets that are certainly confirmed. We restricted the search to cases where the rotation period of the host star is known. We obtained a (non-exhaustive) list of exoplanets for which the ratio of the orbital period ($P_{\rm orbit}$) to the stellar rotation period ($P_{\rm rot}$) can be calculated. Fig.~\ref{fig:porb_prot} shows those ratios in the range from 0 to 1, as well as the planet eccentricities. Highlighted in red are the exoplanets for which the ratio $P_{\rm orbit}/P_{\rm rot}$ is close to either 0.4
or 0.75, i.e. the ratios expected from the $m=2$ and $m=3$ sectoral r~modes \citep[see][]{2000ApJ...531..415H,2001ApJ...551.1107F,2006PASP..118.1690M,2010MNRAS.408.1666S,2013A&A...556A.110B,2013A&A...556A.126A,2014ApJ...791..114W,2015MNRAS.452.2745S,2015ApJ...803....8H,2015MNRAS.452.2745S,2016A&A...595A..12S,2018A&A...619A...1B}.
Large eccentricities correspond to RV signals that are not sinusoidal, unlike r-mode linear oscillations that are expected to be sinusoidal. Since HD~38529~b has an eccentricity of 0.248, it is unlikely to be an r-mode false positive. Note that GJ~163, HD~130322, and GJ~667~C were observed in RV over short periods, typically less than 1000 days (corresponding to 9-13 stellar rotations); these are perhaps the  most likely candidates for r-mode oscillations since r modes may remain coherent over such periods of observation. The signals for GJ~163~c, GJ~832~c and GJ~667~C~f have RV semi-amplitudes of 3~m/s or below, while HD~130322~b and 55~Cnc~b have RV semi-amplitudes of 112~m/s and 71~m/s, which is much above the solar r-mode amplitudes. 

The amplitude of the RV variations associated with active regions are of the order of several m/s in the Sun and slowly rotating stars \citep[e.g.,][]{Haywoodetal16,Lanzaetal16,Lanzaetal18}. Several methods have been developed to correct for such activity-induced variations based on different techniques and correlations with activity indicators \citep[e.g.,][]{Dumusqueetal17}. In the best cases, they can be applied to uncover Keplerian signals with amplitudes of several tens of cm/s in long time-series of late-type stars with very low levels of activity  \citep[and convective noise, e.g.,][]{Dumusqueetal12}. It is precisely in the 10~cm/s regime, which is becoming accessible thanks to the increasing stability and accuracy of stellar speedometers such as HARPS and ESPRESSO, that solar-like r~modes could become a source of false detections. They can be identified by their characteristic frequencies provided that the stellar rotation period is known. We note that the relationship between r-mode amplitudes and activity-cycle phase is not straightforward for the Sun (Z.-C. Liang, priv. comm.).  Present methods to correct for the effects of stellar activity in RV timeseries are not yet advanced enough as to reach the precision of 10~cm/s, thus making it even more important to understand all the source of RV variations at that level, including the effects of the r modes.

A solar telescope has been built to feed the high-accuracy spectrograph HARPS-N to make  measurements of the solar RV from its spectrum integrated over the solar disc. The system collects up to $\sim 40-60$ measurements per clear day, reaching a precision of $\sim 35-40$ cm/s on the single measurement  \citep{Dumusqueetal15,Phillipsetal16}. A preview of the results has been published by, e.g., \citet{MortierCollierCameron17}. In the lower panel of their Fig.~2, the stacked periodogram of a series of measurements covering approximately 300 days shows signals with {\it synodic} periods around 19 and 9 days, which are rather close to the expected periods of the $m=2$ and $m=3$ sectoral r~modes.   If these oscillation periods are indeed due to r modes, then they should also be present in the BiSON and GOLF/SOHO disk-integrated velocity observations covering multiple decades (at least at times when $B_0\neq0$), as well as in other spatially-resolved Doppler observations.
Furthermore, we conjecture that r modes may affect the RV of the Sun or  late-type stars also through an indirect way, that is, by modifying the distribution of the surface magnetic fields through their velocity field because magnetic flux tubes are frozen to the photospheric plasma. We expect the magnetic fields to be concentrated in regions of higher vorticity \citep[e.g.,][]{Balmacedaetal10}, thus producing a local quenching of the convective spectral line blueshifts that can lead to a perturbation of the disc-integrated RV \citep[e.g.,][]{Meunieretal10} with the temporal and spatial periodicities of the underlying r modes. 

\section{Conclusions}
The recent discovery of r~modes in the Sun by \citet{Loptienetal18} prompted this investigation of their possible signature in the  RV measurements of sun-like and late-type stars, including those that are targets for exoplanet search by means of the radial velocity technique. 

Under reasonable assumptions, we find that the lowest-$m$ r~modes may produce disc-integrated RV oscillations with amplitudes up to several tens of cm/s. Their periods are  comparable with the stellar rotation period, but different from it and its harmonics, unlike the modulations produced by stellar activity. False planetary detections produced by r-mode oscillations, if any, should be identifiable provided that the rotation period of the star is known.  We expect putative orbital periods of the order of 10-20 days and amplitudes corresponding to massed of 0.1-0.2 Earth masses, if we consider the low-amplitude r modes excited in the Sun. However, larger amplitudes could be possible in other stars that might put into question some of the candidates recently proposed around some M dwarfs (cf. Sect.~\ref{discussion}). 

\begin{acknowledgements}
The authors wish to thank an anonymous referee for a careful reading of their manuscript and valuable comments. They gratefully acknowledge support from the International Space Science Institute in Bern in the premises of which this work was initiated in July 2018.  AFL acknowledges support from the Italian Ministry's {\it Progetti Premiali} to INAF (Premiale Frontiera). TZ was supported by the Austrian Science Fund (FWF) project 30695-N27 and by Georgian Shota Rustaveli National Science Foundation project 217146. LG acknowledges support from the Max Planck Society through a grant on PLATO science. KR is a member of the International Max Planck Research School for Solar System Science at the University of G\"ottingen. KR contributed to the discussion section, especially Figure~\ref{fig:porb_prot}.
\end{acknowledgements}

\end{document}